\documentclass[reprint,amsmath,amssymb,prl]{revtex4-2}
\usepackage{graphicx}
\usepackage{dcolumn}
\usepackage{bm}
\usepackage{comment}
\usepackage[usenames,dvipsnames,svgnames,table]{xcolor}
\usepackage[capitalise]{cleveref}

\begin{document}


\title{Micromechanical origin of plasticity and hysteresis in nest-like packings}

\author
{Yashraj Bhosale$^{1\ast}$, Nicholas Weiner$^{2\ast}$, Alex Butler$^{2}$, Seung Hyun Kim,$^{1}$, Mattia Gazzola$^{1,3,4\dagger}$, Hunter King$^{2,5\dagger}$\\
	\normalsize{$^{1}$Mechanical Sciences and Engineering, University of Illinois at Urbana-Champaign, Urbana, IL 61801, USA}\\	
	\normalsize{$^{2}$School of Polymer Science and Polymer Engineering, University of Akron, Akron, Ohio 44325, USA}\\
	\normalsize{$^{3}$National Center for Supercomputing Applications, University of Illinois at Urbana-Champaign, Urbana, IL 61801, USA}\\
	\normalsize{$^{4}$Carl R. Woese Institute for Genomic Biology, University of Illinois at Urbana-Champaign, Urbana, IL 61801, USA}\\
	\normalsize{$^{5}$Department of Biology, University of Akron, Akron, Ohio 44325, USA}\\
	\normalsize{$\ast$These two authors contributed equally.}\\
	\normalsize{$^\dagger$To whom correspondence should be addressed: mgazzola@illinois.edu, hking@uakron.edu.}
}


\date{\today}

\begin{abstract}

Disordered packings of unbonded, semiflexible fibers represent a class of materials spanning contexts and scales.  From twig-based bird nests to unwoven textiles, bulk mechanics of disparate systems emerge from the bending of constituent slender elements about impermanent contacts. 
In experimental and computational packings of wooden sticks, we identify prominent features of their response to cyclic oedometric compression: non-linear stiffness, transient plasticity, and eventually repeatable velocity-independent hysteresis. We trace these features to their micromechanic origins, identified in characteristic appearance, disappearance, and displacement of internal contacts. 

\end{abstract}

\maketitle



When a bird forages for nesting material, it may poke or shake a candidate stick before deciding whether or not to add it to the growing nest structure~\cite{hansell}.  
In so doing, it instinctively relates the mechanical properties of the element to its properties \emph{in aggregate}, which are tasked with crucial roles in the protection and development of the bird's unhatched offspring~\cite{collias,deeming}. 
That apparent foresight is particularly surprising given the complexity of the resulting material and our own limited understanding of its emergent mechanics.  What defines this material and how does it depend on its basic ingredients?

By conceptually simplifying the nest to a disordered packing of slender grains, we frame our questions through the lens of granular physics.  Indeed, its solid state follows from the `jamming' of its elements \cite{jam-review}, whose contacts prevent them from flowing around each other, leading to reproducible self-assembly under generic confinement, robustly quantifiable by volume fraction $\phi$ and average contact number $<z>$ \cite{philipse1996packing,fraden,wouterse2009contact,tallinen}. 
A granular metamaterial, the nest's behavior depends both on the base material of its elements and their shape and friction.  Slenderness causes them to pack at low volume fraction and gives rise to greater resistance to flow~\cite{franklin,column-collapse}.  It also introduces bending as an additional mode of accommodating external stresses.  The contact network that propagates stresses through the aggregate is transient to the extent that friction allows elements to slide against one another.  The resulting mechanical responses are rich, but rigorously characterizable, and occupy a space between hard grains~\cite{parafiniuk2018ellipses,parafiniuk2016rods,gravish2012entangled,jaeger2015jamming} and unwoven fabrics \cite{picu2011review,weiner2020mechanics}.

Nest-material softness manifests via two specific mechanical behaviors which we observe in response to cyclic oedometric compressive strain: (1) Transient plasticity during initial cycles, after which non-linear stress-strain curves adopt a repeatable, steady-state shape; (2) Subsequent finite hysteresis independent of sufficiently small loading speeds, indicating a non-viscous mechanism of dissipation.  
Our aim is to quantify these at the grain and bulk scale to illuminate the micromechanical origin of each.


\begin{figure*}[t]
\includegraphics[width=0.8\linewidth]{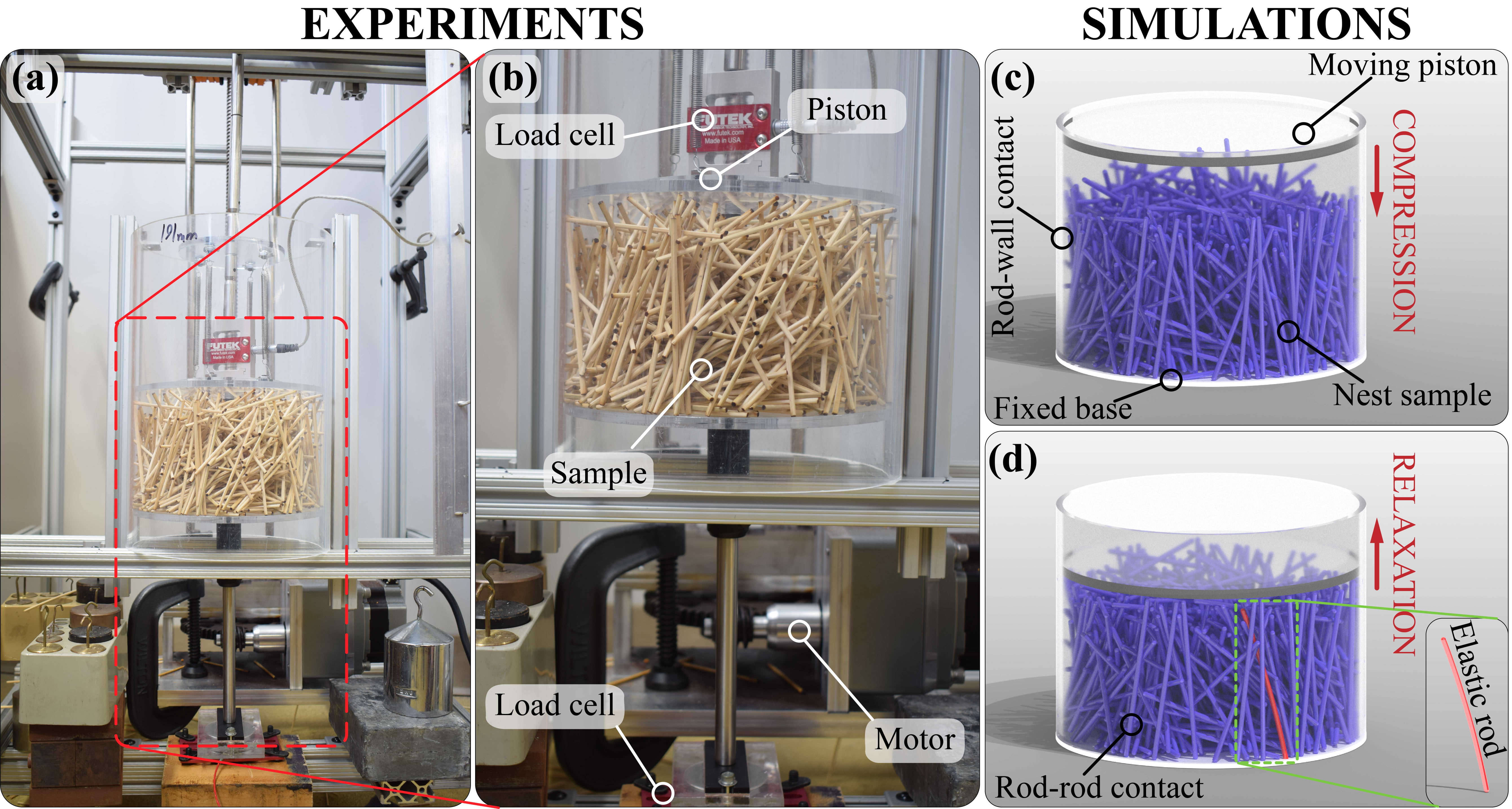}
\caption{\label{fig:setup} Illustration of experimental and computational setups. (a) Overall view of the experimental mechanical compression setup, where the random packing of sticks is bounded by an acrylic cylinder, and compression is performed by a moving top lid driven by a motor. The applied load is then measured using the load cells at the bottom of the cylinder. (b) Zoomed-in view of the experimental setup with the critical parts highlighted. Computational counterpart of the mechanical testing setup, with the random packing sticks shown in (c) relaxed and (d) compressed states. Details regarding both setups can be found in the SI.}
\end{figure*}

\begin{figure*}[t]
\includegraphics[width=0.75\linewidth]{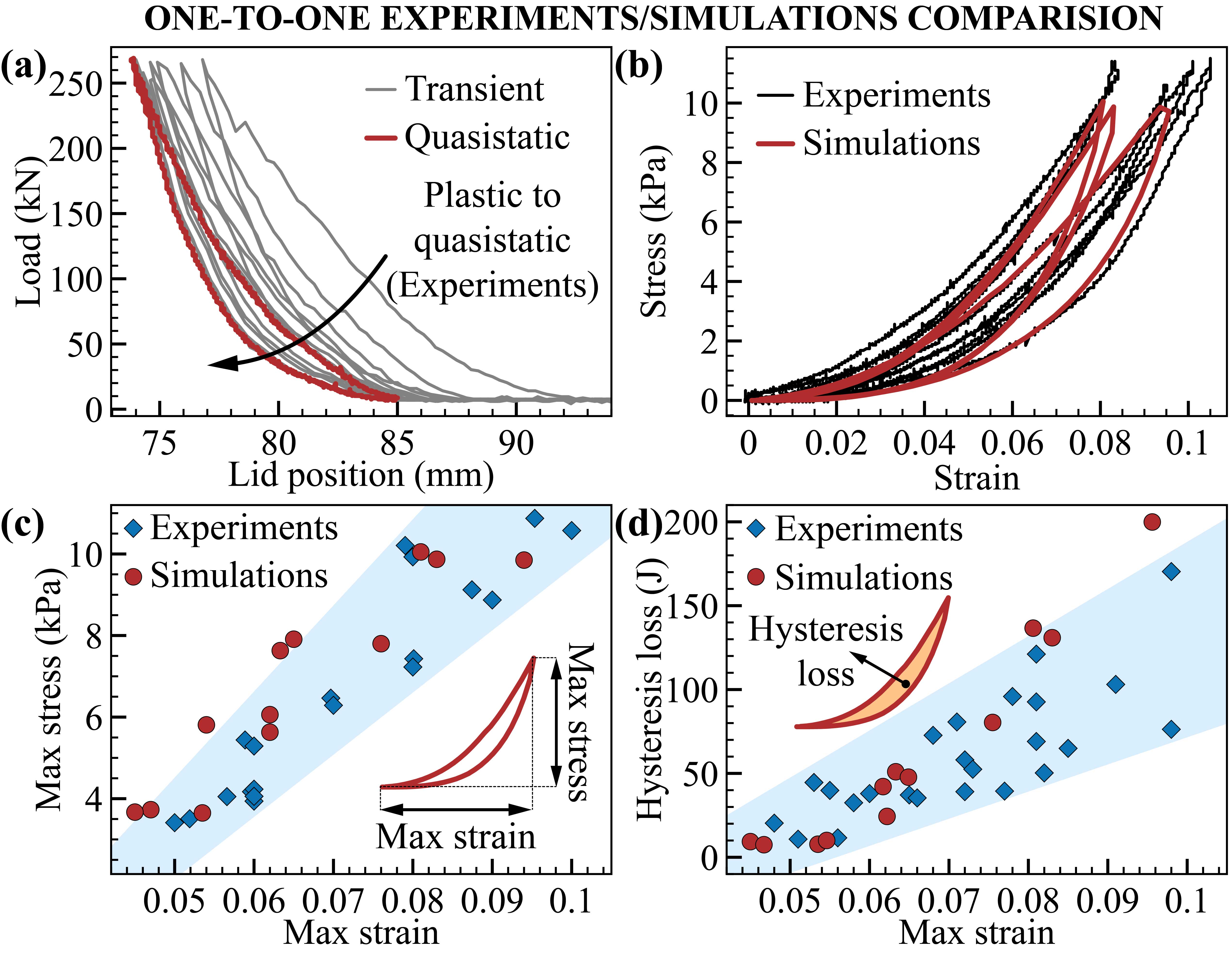}
\caption{\label{fig:validation} Bulk mechanical response characterization and validation. (a) Experimental variation of the packing load with the cylinder lid position, showing initial transience (grey) and final quasi-static response/cycle (red). Comparison between simulations and experiments, for (b) quasistatic stress-strain cycles, (c) max stress vs. max strain, and (d) hysteretic losses. }
\end{figure*}

Random packings of 460 bamboo rods, each of aspect ratio AR=31 (L=76mm and D=2.4mm), are prepared in a cast acrylic cylinder of 140mm diameter by pouring them in a randomized fashion so as to avoid ordering artifacts at the walls.
Oedometric compression cycles are then performed by the custom mechanical tester of \cref{fig:setup}(a,b). The setup consists of a load cell with a stationary bottom plate supporting the sample confined by cylindrical walls. A moving top plate, equipped with a second load cell, is driven by a stepper motor to apply pressure to the sample.  Actuation of the motor, height tracking and measurement from the two load cells are coordinated through a programmable microcontroller.



In a one-to-one computational counterpart, compression and relaxation phases are illustrated in \cref{fig:setup}(c) and (d), respectively. Bamboo sticks (purple) are modelled via the software \textit{Elastica} \cite{Gazzola2018,Zhang2019} as Cosserat rods, employing the same geometry, bending modulus ($B = 12$ GPa, measured with 3-point tester) and static friction coefficient ($\mu_{\textrm{static}} = 0.3$, measured with tilt-angle jig) of experiments. A virtual cylindrical container and a moving piston constrain and cyclically compress the randomly initialised packing. Rod-rod and rod-boundary interactions account for both contact and friction. Upon detecting collision, a repulsive force based on Hertzian contact theory \cite{hertz2021ueber} is applied to the rods to prevent interpenetration. The Hertz-Mindlin model \cite{yan2015discrete,hertz2021ueber,mindlin1949compliance} is employed throughout the system (both among rods and with the boundaries) to capture 'stiction' (slip-stick due to static friction), under the assumption of isotropy. In all numerical studies, parameters are matched to the corresponding experimental values (number of sticks, static friction, bending stiffness, container geometry, compression protocol). Details can be found in the SI.

\begin{figure*}[ht]
\includegraphics[width=0.85\linewidth]{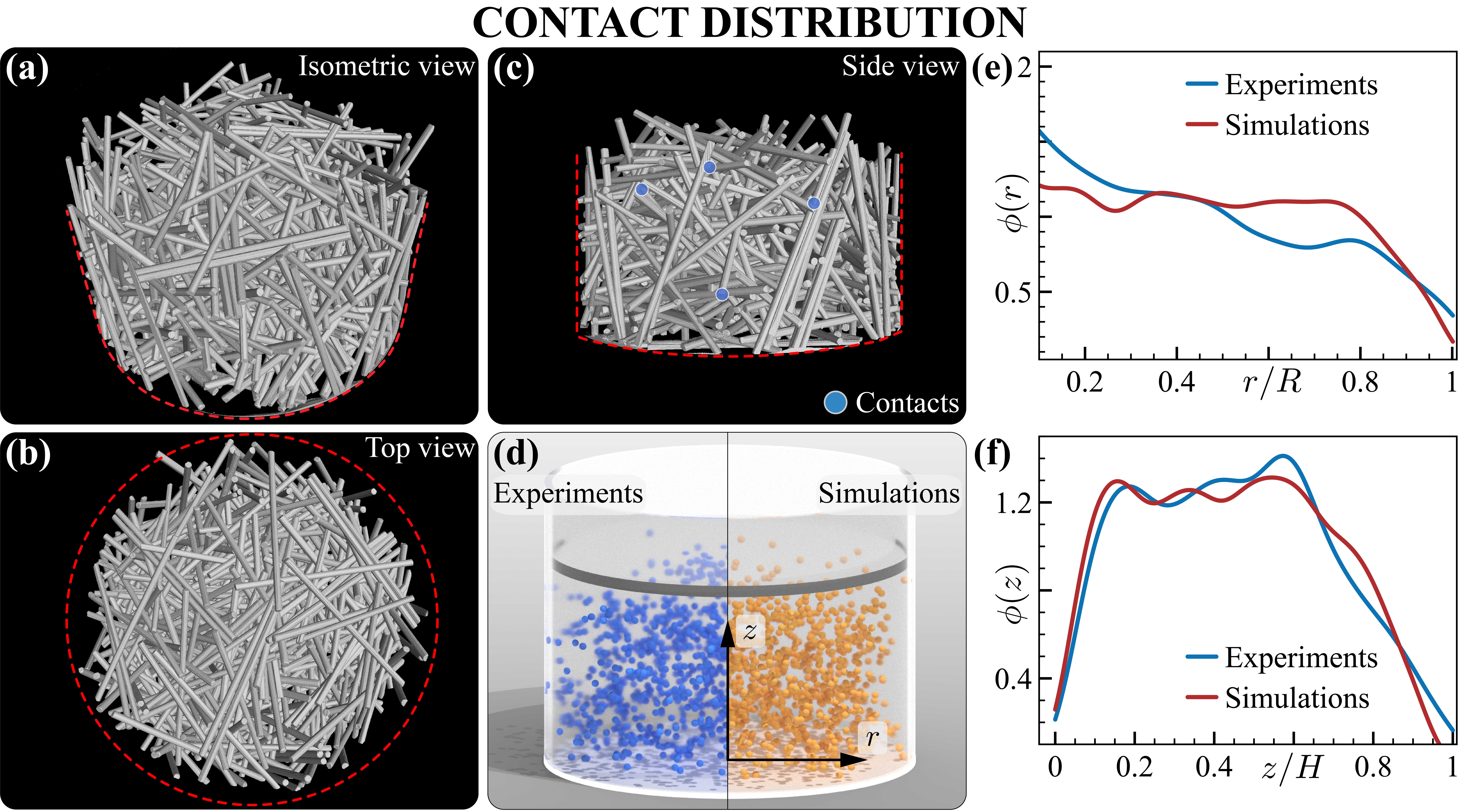}
\caption{\label{fig:contact} Packing internal structure visualisation and contact distribution. The packing skeleton is visualised via CT scan renderings, for which the (a) isometric (b) top and (c) bottom view are presented. (d) Comparison of the spatial 3D distribution of contacts, between simulations and experiments. (e) Comparison of the contact density variation with the radial coordinate, i.e. from the packing center to the cylinder boundary, between simulations and experiments. (f) Comparison of the contact density variation with the axial coordinate, from the bottom to the top of the packing, between simulations and experiments.}
\end{figure*}


With experimental and numerical setups in hand, we proceed with characterizing the nest behavior and validating our simulations. 
\Cref{fig:validation}(a) shows a typical quasistatic loading response for repeated cycles of fixed maximum load. The initial cycles display significant plasticity, as the sample returns to an increasingly compact zero-stress state and the curves move to the left.  After several cycles, the system begins to retrace a consistent `steady-state' loop. In this regime, non-linear loading and unloading curves carve an area which represents energy dissipation no longer associated with plasticity.  This behavior is qualitatively the same across a range of experimental parameters such as stick number and aspect ratio (see SI).
\Cref{fig:validation}(b) presents a comparison of these curves (expressed as stress versus strain) between experiments (black) and simulations (red), for multiple preparations of a single set of sticks (AR=31, stick number=460). 
Experimental and numerical results share the same qualitative features: consistent non-linear shape and pronounced hysteresis.  Their quantitative differences fall within the deviations seen in either set of results.  These deviations appear to correlate with packing fraction, which varies between individual preparations about a robust average ($\phi$=0.134\textpm 0.006). 

To quantitatively compare experimental and computational data over many samples, we extract two parameters from the steady state curves, as in Refs.~\cite{athanassiadis2014particle,parafiniuk2018ellipses}. 
First, an effective stiffness is characterized as the maximum stresses reached for a range of maximum applied strains.  \Cref{fig:validation}(c) shows experimental (blue) and numerical (red) values for max stress as a function of max strain, which agree well within the range spanned by experiments (shaded region). Second, energy dissipated per cycle, calculated by taking the integral between the loading and unloading curves, is presented as a function of max strain, for experiments (blue) and simulations (red), in \cref{fig:validation}(d). Again, simulations agree well with experiments, demonstrating robustness and accuracy in capturing the nest's emergent bulk behavior.  We note that the observed stress-strain curves strongly resemble a viscoelastic response, but emphasize that neither their shape nor the size of hysteretic losses is dependent on strain rate (as demonstrated for 3 orders of magnitude variation in speed, see SI). 
This implies that a fundamentally different, quasi-static mechanism is responsible for the dissipation, one which depends on a cyclic asymmetry in the system's instantaneous configuration, rather than its dynamics.

\begin{figure*}[ht]
\includegraphics[width=0.85\linewidth]{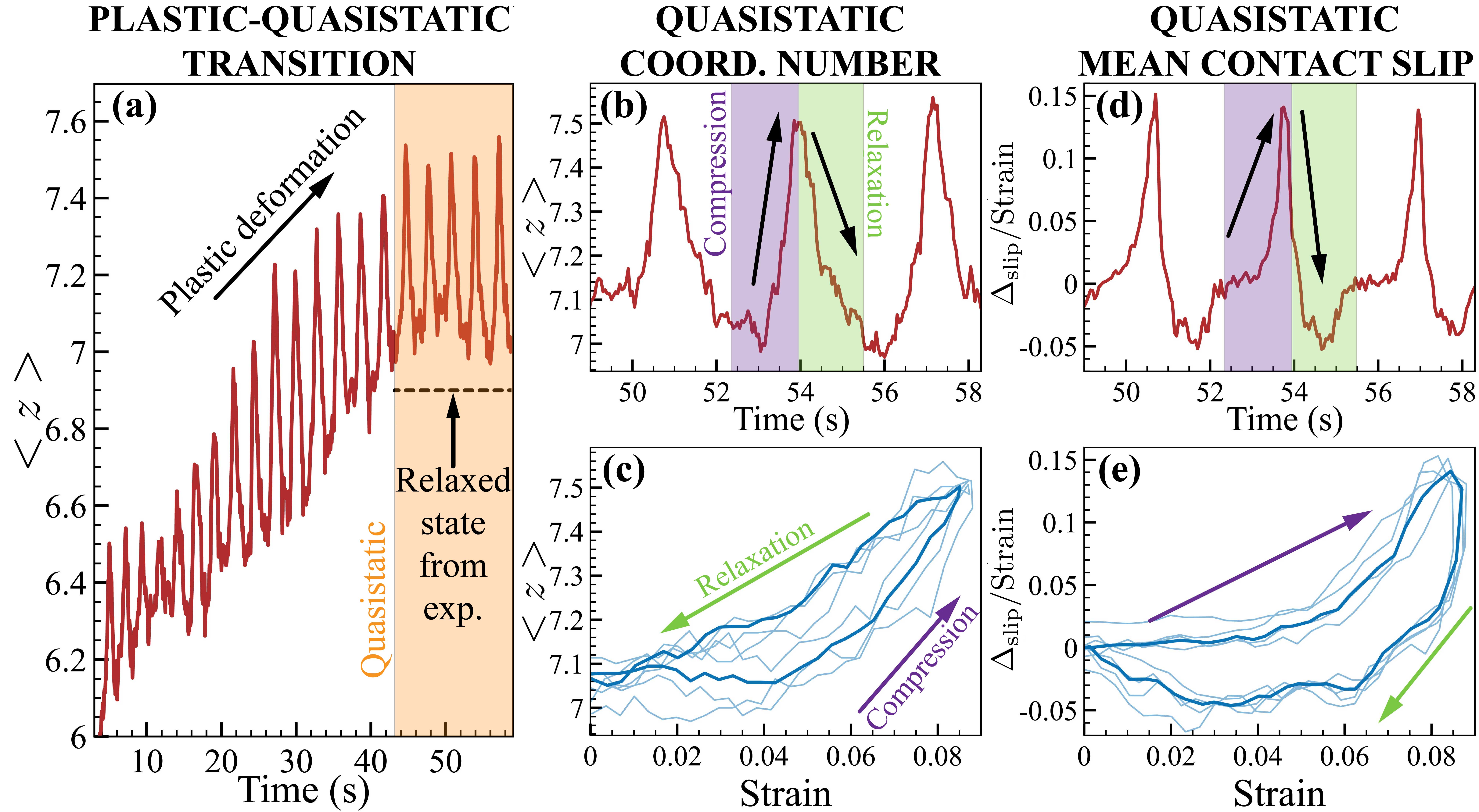}
\caption{\label{fig:micromechanics} Micro-mechanical response. (a) Coordination number vs. time for the initial plastic cycles and the quasistatic regime (highlighted in orange). Coordination number variation in the quasistatic regime, against (b) time and (c) strain. Variation of averaged contact slippage relative to the cycle strain in the quasistatic regime, plotted against (d) time and (e) strain. }
\end{figure*}

We therefore turn our attention to the nest's internal structure. Three dimensional configurations of unloaded samples, which had been previously compressed beyond transient plasticity, were acquired by computer assisted tomography (Nikon XT H 230). Visualizations using an isometric, top and side view are seen in \cref{fig:contact}(a), (b) and (c), respectively. 
Individual rod cross-sections were segmented by applying a watershed transform~\cite{roerdink2000watershed} to the distance map of binarized cross-sectional images, following a procedure similar to Ref.~\cite{tallinen}.  Location of individual contacts between rods were identified by dilating labeled rods, creating clouds of overlapping voxels, and then segmenting those clouds as individual contacts (see SI). A resulting 3D reconstruction of contacts is shown in \cref{fig:contact}(d), where it is compared with that of simulations. 

Non-dimensional spatial densities of inter-element contact points, as functions of radial distance from sample center and vertical distance from cylinder base, are compared in \Cref{fig:contact}(e) and (f), respectively. As seen in \cref{fig:contact}(e), contact density is roughly uniform within a central core occupying $\sim$80\% of the packing, before dropping near the confining wall, where fewer contacts are observed. \Cref{fig:contact}(f) shows a similar `boundary layer' effect, where the contact density decreases both at the top and bottom from an approximately constant distribution near the sample center.
We emphasize the remarkable agreement with simulations, which are therefore shown to capture both macro- and micro-mechanical salient features. 

Then, in search of an explanation for the observed nest emergent properties, we take advantage of simulations to correlate macroscopic behavior to microscopic contact formation, displacement, and disappearance.
For this purpose, we track contacts during loading cycles, and characterize them via two diagnostics. First is the average number of contacts per rod in the system, also known as the coordination number $<z>$. Second is the relative motion of the contacts along their corresponding rods, captured via the average slippage over a cycle.

\Cref{fig:micromechanics}(a) shows the evolution of the coordination number as a function of time. As the system rearranges, cycle after cycle, the coordination number increases until it reaches a steady state regime, implying a maximally packed state.  
In this regime, the instantaneous coordination number oscillates from $\sim 7$ in relaxed mode to $\sim 7.5$ in compressed mode. This is consistent with our CT scan analysis, where a coordination number of $\sim$ 6.9 is recovered for a maximally packed preparation in relaxed mode. 
Previous studies~\cite{freeman2019boundary,blouwolff2006coordination} report values approaching the theoretical estimate $<z>$ = 10 for packed rods of similarly high aspect ratio~\cite{philipse1996packing}.  We attribute the discrepancy to the boundary effects discussed above. Indeed, if we compute a coordination number from the more uniform, central 80\% portion of the system, we find $<z>$ = 9.4, much closer to the theoretical prediction.

We now zoom into the identified maximally packed regime, to gain further insight into the micromechanics at play.
\Cref{fig:micromechanics}(b-c) shows the evolution of the coordination number during quasistatic cycles, as a function of time and strain, respectively.  In both cases, an asymmetric behavior in the creation and disappearance of contacts is registered in loading and unloading strokes.  
The creation of contacts during compression necessarily implies increased connectivity of the contact network which propagates stresses, as well as a decrease in the average distance between contacts about which sticks are compelled to bend.  Both factors favor increased stiffness with increasing load, which is consistent with the observed nonlinearity in the stress-strain curves (\Cref{fig:validation}b).
The fact that the release of contacts during the unloading stroke is not symmetric in strain is instead consistent with the different slopes, resulting in the observed, characteristic hysteresis.  Although insightful, these observations alone do not explain the mechanistic origins of the asymmetry.


For that, we further consider the mean slippage of the contacts along their corresponding rods, non-dimensionalized by the cycle strain, as function of time and strain itself. As can be seen in \cref{fig:micromechanics}(d-e), the contacts slide along the length of the rods in one direction during loading and then return to their initial position during unloading, in a repeating pattern. The pattern itself is asymmetric with respect to strain, such that the onset of motion occurs at a higher strain during compression than it does during relaxation.

This mechanism is consistent with a `ratcheting' role of static friction.  Lateral forces due to internal stress must overcome static friction at individual contact locations, causing a phase lag in their motion with respect to system strain, which in turn manifests as a lag in the effective stiffness of the system, leading to hysteretic response.  Energy must still dissipate as heat, but the mechanism is inherently non-viscous as it does not depend on strain rate, as previously mentioned.
This picture agrees with reversible slippage speculated without direct observation in the context of other granular systems of spheres and ellipsoids~\cite{khalili2017numerical, parafiniuk2018ellipses}, and is conceptually similar to a proposed pure-elastic dissipation mechanism in elastomers~\cite{farris1}.  

In conclusion, our study establishes the systematic experimental and computational characterization of a class of metamaterials, inspired by bird nest, at the interface between grains and textile fibers.  Our analysis shows that such nest packings exhibit non-linear stiffness and quasi-static hysteresis.  These robustly reproducible features are found to naturally emerge from the rearrangement of the frictional contact network in response to external confinement, demonstrating a granular dissipation mechanism previously only speculated and fundamentally distinct from viscous processes. These results serve as a starting point toward the characterization of nest-like materials in general and practical settings.
Deeper understanding of the relationship between aggregate mechanics and element properties, as intuited by the bird engineer, may lend insight into the phenomenology of diverse systems sharing similar ingredients, including unwoven fibers, unbonded collagen and fungal mycelium networks~\cite{munster2013strain,islam2017morphology,poquillon2005fibers}. 
Further, these lightweight, conformal and reusable structures may prove useful as architectural and engineering materials alike~\cite{dierichs2016architecture}.


\begin{acknowledgments}
We thank L. Mahadevan for inspiring interest in the problem. This work was financially supported by the National Science Foundation (NSF) (Award Nos. ENG-1825440, 1825924). 
The authors acknowledge support by the Blue Waters project (OCI- 0725070, ACI-1238993), a joint effort of the University of Illinois at Urbana-Champaign and its National Center for Supercomputing Applications. This work used the Extreme Science and Engineering Discovery Environment (XSEDE) Stampede2, supported by National Science Foundation grant number ACI-1548562, at the Texas Advanced Computing Center (TACC) through allocation TG-MCB190004. 
\end{acknowledgments}

\bibliography{apssamp}

\end{document}